# Implications of Neutron Star Mergers for Extraterrestrial Civilizations

The economy and fate of extraterrestrial civilizations should depend on the abundance of gold and uranium, made in neutron star mergers

———

By Abraham Loeb on March 8, 2018

Astronomers have recently determined that rare elements such as gold and uranium, are produced as a result of rapid capture of free neutrons during the merger of two neutron stars. Neutron stars are the densest stars known, having the size of a city (12 kilometers) and up to twice the mass of the Sun, with the density of an atomic nucleus. A teaspoon of neutron star material weighs a trillion kilograms, as much as a tall mountain on Earth.

Last summer the LIGO experiment detected ripples in spacetime (so-called gravitational waves) that were produced by the merger of two neutron stars in a distant galaxy. The light emitted after the merger provided evidence that a small fraction of the neutron-rich matter was ejected into space and transformed to elements such as gold and uranium. The expected rate of such mergers can account for the average abundance of these elements in galaxies.

But neutron star collisions are extremely rare and so the abundance of gold or uranium is expected to vary considerably in space and time within any given galaxy, as long as subsequent mixing processes are incomplete. Indeed, spectroscopic studies of Galactic stars reveal a large spread in their abundance of Europium, a related heavy element. Under these circumstances, one cannot help but wonder how the history of extragalactic civilizations would be shaped by their distance from the nearest neutron star merger.

Imagine an extraterrestrial (ET) civilization on the surface of a planet that is deficient in uranium relative to Earth. Such a civilization will be less likely to develop nuclear weapons and to annihilate itself through a nuclear war. A lower abundance of gold would make this element more precious and have a major impact on the ET economy. Hence, both the economy and longevity of an ET civilization will depend in part on how far it is from its nearest neutron star merger. For safety reasons, we should avoid a conflict with a Uranium-rich civilization that was born near a neutron star merger site.

There is no doubt that the cosmic neighborhood matters in shaping the life of civilizations. For example, Proxima b is the planet orbiting the nearest star to the Sun, Proxima Centauri. Even though the planet is merely 4.2 light years away from us, any form of life on Proxima b must experience a very different life than ours. Since the host star is faint (having 12% of the mass and 0.2% of the luminosity of the Sun), the habitable zone around it - where Proxima b resides - is twenty times closer than the Earth is from the Sun. Given this proximity, the planet is likely to be tidally locked, showing the same face to its star as it orbits around the star every 11.2 (instead of 365) days. Proxima b is therefore likely to possess a hot permanent day side and a cold permanent night side. If the planet is covered by an atmosphere, it would exhibit strong winds, and its weather would fluctuate due to the variations in the powerful stellar wind and stellar flares at its close distance to Proxima Centauri. This could motivate civilizations to develop protective technologies around dwarf stars. For example, by covering the permanent day side with photovoltaic cells, ETs could transfer electric power to heat and illuminate the otherwise cold and dark night side of their planet. Similarly, an artificially constructed magnetic loop could shield the planet from energetic particles in stellar flares.

The quality of Galactic life should also depend on the distance from the center of the Milky Way. If we were ten times closer to the Galactic center, then bright flares of the supermassive black hole there, SgrA* (which weighs four million solar masses), would have stripped the Earth's atmosphere. On the positive side, such flares could have also transformed mini-Neptunes into Earths, increasing the abundance of habitable planets.

In the distant future, once we will develop the technology to transport our civilization to another planet, we will be faced with the dilemma of which planet to inhabit. My twelve-year old daughter told me that she would buy two houses on Proxima b, one on the permanent night side for her to sleep in and the second on the permanent sunset strip where she will take her vacations.

And she will likely not be alone. Real estate values must peak on the permanent sunset strip of tidally-locked exoplanets. If you do not believe me, check it out with the nearest Galactic real estate agent.

## References

Abbott, B. P., et al. 2017, "GW170817: Observation of Gravitational Waves from a Binary Neutron Star Inspiral", Phys. Rev. Lett **119**, 161101.

Anglada-Escude', G. et al. 2016, "A Terrestrial Planet Candidate in a Temperate Orbit Around Proxima Centauri", Nature **536**, 437-440.

Chen, H., Forbes, J. C., & Loeb, A. 2018, "Habitable Evaporated Cores and the Occurrence of Panspermia Near the Galactic Center", ApJ **855**, L1-L8.

Chornock, R. et al. 2017, "The Electromagnetic Counterpart of the Binary Neutron Star Merger LIGO/Virgo GW170817. IV. Detection of Near-infrared Signatures of r-process Nucleosynthesis with Gemini-South", ApJL **848**, L19-26.

Chruslinska, M. et al. 2017, "Double Neutron Stars: Merger Rates Revisited", https://arxiv.org/abs/1708.07885

Delgado, M. E. et al. 2017, "Chemical Abundances of 1111 FGK Stars from the HARPS GTO Plant Search Program. II. Cu, Zn, Sr, Y, Zr, Ba, Ce, Nd, and Eu", Astronomy & Astrophysics **606**, 94-114.

Forbes, J. C. & Loeb, A. 2017, "Evaporation of Planetary Atmospheres Due to XUV Illumination by Quasars", https://arxiv.org/abs/1705.06741

Kasen, D. et al. 2017, "Origin of the Heavy Elements in Binary Neutron Star Mergers from a Gravitational Wave Event", Nature **551**, 80-84.

Lingam, M. & Loeb, A. 2017, "Natural and Artificial Spectral Edges on Exoplanets", MNRAS **470**, L82-86.

Lingam, M. & Loeb, A. 2017, "Risks for Life on Habitable Planets from Superflares of Their Host Stars", ApJ **848**, 41-54.

Lingam, M. & Loeb, A. 2017, "Impact and mitigation strategy for future solar flares", https://arxiv.org/abs/1709.05348

Loeb, A. 2017, "Searching for Life Among the Stars", PEN: Science & Technology **24**, (Profile 1-4).

## ABOUT THE AUTHOR

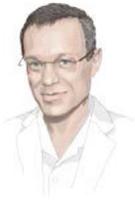

**Abraham Loeb**

Abraham Loeb is chair of the astronomy department at Harvard University, founding director of Harvard's Black Hole Initiative and director of the Institute for Theory and Computation at the Harvard-Smithsonian Center for Astrophysics. He also chairs the advisory board for the Breakthrough Starshot project.

Credit: Nick Higgins